\documentstyle[11pt]{article}
 \let\msk=\medskip  \let\qd=\quad
\let\a=\alpha \let\be=\beta \let\g=\gamma \let\de=\delta
\let\ep=\varepsilon  \let\h=\eta 
\let\dh=\vartheta  \let\la=\lambda 
   \let\si=\sigma
\let\om=\omega \let\ps=\psi
 \let\Ph=\phi \let\PH=\Phi \let\Ps=\Psi
\let\Om=\Omega  \let\TH=\Theta
\let\La=\Lambda \let\Ga=\Gamma \let\De=\Delta
\def\0#1#2{\frac{#1}{#2}}
\def\s0#1#2{\mbox{\small{$\frac{#1}{#2}$}}}
\def\5{\bar }  \def\6{\partial } \def\7{\hat } \def\4{\tilde }
\let\LRA=\Leftrightarrow \let\then=\Rightarrow
\def\sl2c{$sl(2,{\bf C})$} \def\SL2c{$SL(2,{\bf C})$}
\def\nin{\in\hspace{-.85em}/\ }
\let\nn=\nonumber
\def\bea{\begin{eqnarray}} \def\eea{\end{eqnarray}}
\def\beann{\begin{eqnarray*}} \def\eeann{\end{eqnarray*}}
\def\beq{\begin{equation}} \def\eeq{\end{equation}}
\def\ba{\begin{array}} \def\ea{\end{array}}
  \def\cg{{\cal G}}
\def\ca{{\cal A}} \def\cd{{\cal D}} 
 \def\cN{{\cal N}}
\def\cF{{\cal F}}  \def\cH{{\cal H}}
\def\cW{{\cal W}} \def\cP{{\cal P}} 
\def\SS{{\cal J}} 
\def\da{{\dot{\a}}} \def\dbe{{\dot{\be}}}
\def\dg{{\dot{\g}}} 
\def\ua{{\underline{\a}}} \def\ube{{\underline{\be}}}
\def\ug{{\underline{\g}}} 
\def\on{{\overline{n}}}
\def\dD{{(w)}} \def\dR{{(r)}}
\def\T#1#2#3{{T_{#1#2}}^{#3}} 
\def\F#1#2#3{{F_{#1#2}}^{#3}} \def\G#1#2#3{{G_{#1#2}}^{#3}}
\def\Gg#1#2#3{{g_{#1#2}}^{#3}} \def\f#1#2#3{{f_{#1#2}}^{#3}}

\def\e#1#2{{e_{#1}}^{#2}} \def\E#1#2{{E_{#1}}^{#2}}
\def\cE#1#2{{{\cal A}_{#1}}^{#2}}
\def\o#1#2{{\om_{#1}}^{#2}}
\def\cA#1#2{{{\cal A}_{#1}}^{#2}} \def\p#1#2{{\ps_{#1}}^{#2}}
\def\dps#1#2{\5\ps_{#1}{}^{#2}}
\def\A#1#2{{A_{#1}}^{#2}}

\def\csum#1#2{\sum_{#1}\hspace{-1.#2em}\circ\ \ \ }
\def\osuo#1#2#3#4{{\sigma^{#1#2}}_{#3}{}^{#4}}
\def\osuu#1#2#3#4{{\sigma^{#1#2}}_{#3#4}}
\def\usuo#1#2#3#4{{\sigma_{#1#2\, #3}}^{#4}}

\def\obsuu#1#2#3#4{\5\sigma^{#1#2}{}_{#3#4}}
\def\ubsou#1#2#3#4{\5\sigma_{#1#2}{}^{#3}{}_{#4}}
\def\so#1#2#3{\sigma^{#1}{}_{#2#3}}

\newcommand{\mysection}[1]{\section{#1}\setcounter{equation}{0}}
\def\Gl#1{(\ref{#1})}
\begin{document}
{\pagestyle{empty}\hspace*{\fill} NIKHEF-H 93--12\\
\hspace*{\fill} ITP--UH 07/93\vspace*{1cm}\\
\begin{center}{\Large {\bf Anomaly candidates and invariants of D=4, N=1
supergravity theories}}\vspace{2cm}

{\renewcommand{\thefootnote}{\fnsymbol{footnote}}
Friedemann Brandt\footnote{Supported by Deutsche Forschungsgemeinschaft}}
\setcounter{footnote}{0}\vspace{.5cm}

Institut f\"ur Theoretische Physik, Universit\"at Hannover,\\
Appelstra\ss e 2, W--3000 Hannover 1, Germany\\
and\\
NIKHEF Section H, Kruislaan 409, Postbus 41882,\\
1009 DB Amsterdam, The Netherlands\end{center}\vspace{2cm}

\begin{abstract}
All anomaly candidates and the form of the most general invariant local
action are given for old and new minimal supergravity,
including the cases where additional Yang--Mills and chiral matter multiplets
are present. Furthermore nonminimal supergravity is discussed. In this case
local supersymmetry itself may be anomalous and some of the corresponding
anomaly candidates are given explicitly. The results are obtained
by solving the descent equations which contain the consistency
equation satisfied by integrands of anomalies and invariant actions.
\end{abstract}\newpage}

\setcounter{page}{1}
\mysection{Introduction}

For a large class of gauge theories one can construct
a nilpotent BRS operator $s$ and use it to characterize
invariant actions and anomaly candidates as nontrivially
BRS-invariant local functionals of the fields \cite{brs}, \cite{bau},
i.e. as solutions $\cW^g$ of the so-called consistency equation
\beq s\, \cW^g=0,\qd \cW^g\neq s\, \Ga^{g-1}
                                                        \label{1}\eeq
where $\cW^g$, $\Ga^{g-1}$ are local functionals of the fields whose
superscript denotes their ghost number.
Invariant actions are solutions $\cW^0$ with ghost number 0,
anomaly candidates are solutions $\cW^1$ with ghost number 1
(in the case $g=1$ \Gl1
contains the consistency conditions which anomalies have to satisfy
and which have been first derived for Yang--Mills theories \cite{wz}).

The present paper contains results of an investigation of \Gl1 for
old minimal supergravity \cite{old}, new minimal supergravity \cite{new}
and nonminimal supergravity \cite{non}, \cite{ggmw}
where in all cases not only pure supergravity
is considered but its coupling to Yang--Mills
multiplets and chiral matter multiplets is included. I remark that
the results for old and new minimal supergravity are complete
i.e. in these cases
there are no solutions $\cW^0$ and $\cW^1$ of \Gl1 apart from those
given here (modulo trivial solutions $s\Ga^{-1}$ resp. $s\Ga^{0}$)\footnote{The
results given for nonminimal supergravity are incomplete.}. However
it is not the subject of this paper to prove this statement
since the proof can be performed in more generality \cite{dok}.
Instead the present paper spells out the general results
of \cite{dok} explicitly for the supergravity theories mentioned above.
This includes the solution of the
so-called descent equations which follow from \Gl1 and read
\beq 0<p\leq 4:\qd s\,\om_p^{g+4-p}+d\,\om_{p-1}^{g+5-p}=0,\qd
                   s\,\om_0^{g+4}=0
                                                         \label{d1}\eeq
where $d=dx^m\6_m$ is the exterior derivative,
$\om_p^{g+4-p}$ denotes a local $p$--form with ghost number ${g+4-p}$
and $\om_4^g$ is the integrand of the solution of \Gl1:
\beq \cW^g=\int \om_4^{g}.
                                                         \label{d2}\eeq
In a more compact form \Gl{d1} reads
\beq \4s\, \4\om^{g+4}=0,\qd \4s=s+d
                                                         \label{d3}\eeq
where $\4\om^{g+4}$ denotes the formal sum of the forms $\om_p^{g+4-p}$
\beq  \4\om^{g+4}=\sum_{p=0}^4\om_p^{g+4-p}.
                                                         \label{d3a}\eeq
The discussion of \Gl{d1} resp. \Gl{d3} will give insight into
the method used to solved \Gl1. Namely in fact the solutions of
\Gl{d3} have been calculated first and then the solutions of \Gl1 have been
obtained from them via \Gl{d2}. This method takes advantage of the fact
that, as one can prove for a large class of gauge theories \cite{ten},
each solution of \Gl{d3} can be written---up to trivial contributions of the
form $\4s\4\om^{g+3}$---entirely in terms of tensor fields and certain
variables which generalize the ordinary connection forms.
This holds in particular for the supergravity theories discussed here
where the variables which generalize the connection forms
are given in \Gl{t15} and \Gl{n9} and the set of
tensor fields is given in \Gl{PH} where $\{\Ps^l\}$ in the case of
minimal supergravity is given by \Gl{Ps} and in the case of
nonminimal supergravity by the fields listed in \Gl{nPs} and the
respective complex conjugated fields:
\beq \4\om^{g+4}=\4\om^{g+4}(\4\xi^A,\4C^I,\4Q,\PH^r).\label{d3b}\eeq

The paper is organized as follows: In section \ref{brs} the
BRS operator for supergravity is given
(the BRS transformations given in this paper agree with those
derived in \cite{bbg}), section \ref{res} recalls
the basic features and the field content of old, new and
nonminimal supergravity and contains the
solutions of \Gl1 and \Gl{d3} for $g=0,1$, and in section \ref{ver} it is
verified that $\4s\4\om^{g+4}=0$ holds for the expressions given
in section \ref{res} for $\4\om^{g+4}$. Finally a short
conclusion is given, followed by three appendices.

The conventions concerning grading, complex conjugation,
$\si$--matrices etc. agree with those used in \cite{glusy} and as there
I use exclusively the component formalismus, i.e. none of the fields
appearing in this paper denotes a superfield.

\mysection{BRS operator}\label{brs}

The BRS operator for supergravity is conveniently constructed by means of the
algebra
\beq [\cd_A,\cd_B\}=-\T ABC\cd_C-\F ABI\de_I,\qd
[\de_I,\cd_A]=-\Gg IAB\cd_B,\qd
[\de_I,\de_J]=\f IJK\de_K                                      \label{s5}\eeq
which is realized on tensor fields.
$\{\cd_A\}$ consists of the covariant derivatives $\cd_a$ and
supersymmetry transformations $\cd_\a$, $\5\cd_\da=(\cd_\a)^*$. $\{\de_I\}$
denotes a real basis of the Liealgebra $\cg$ of the structure group which in
general is the direct sum of the Lorentz algebra $\cg_L$, a further semisimple
Liealgebra $\cg_s$ and abelian part $\cg_a$. The Lorentz generators are
denoted by $l_{ab}=-l_{ba}$, the elements of $\cg_s+\cg_a$ by $\de_i$.
\[ \{\cd_A\}=\{\cd_a,\cd_\a,\5\cd_\da\},\qd \{\de_I\}=\{\de_i,l_{ab}:a>b\},
\qd \de_i\in\cg_s+\cg_a,\qd l_{ab}\in\cg_L.\]
In \Gl{s5} the structure constants of $\cg$ are denoted by $\f IJK$ and
the $\Gg IAB$ are the entries of the matrices $g_I$ which represent $\cg$
on the $\cd_A$. The only nonvanishing $\Gg IAB$ occur for $\de_I\in\{
l_{ab},\de_\dD,\de_\dR\}$ where $\de_\dD$ and $\de_\dR$ denote the generators
of Weyl transformations and of $U(1)$ transformations called
R-transformations:
\beq\ba{lll} [l_{ab},\cd_c]=\h_{bc}\cd_a-\h_{ac}\cd_b,&
[l_{ab},\cd_\a]=-\usuo ab\a\be\cd_\be,&
[l_{ab},\5\cd_\da]=\ubsou ab\dbe\da\5\cd_\dbe,\\ \mbox{}
[\de_\dR,\cd_a]=0,&
[\de_\dR,\cd_\a]=-i\,\cd_\a,&
[\de_\dR,\5\cd_\da]=i\, \5\cd_\da,\\ \mbox{}
[\de_\dD,\cd_a]=-\cd_a,&
[\de_\dD,\cd_\a]=-\s0 12\, \cd_\a,&
[\de_\dD,\5\cd_\da]=-\s0 12\, \5\cd_\da.
\ea\label{s7}\eeq
The gauge fields corresponding to $\cd_a$, $\cd_\a$, $\5\cd_\da$, $l_{ab}$ and
$\de_i$ are the components $\e ma$, $\p m\a$, $\dps m\da$, $\o m{ab}$, $\A mi$
of the vierbein, the gravitino and its complex conjugate, the spin connection
and the gauge fields of $\cg_s+\cg_a$ respectively. They are used to
define the covariant derivatives of a tensor field $\PH$ as in \cite{des} by
\beq \cd_a\, \PH=\E am(\6_m-\p m\ua\cd_\ua-\cA mI\de_I)\, \PH\
 \LRA\ \6_m\, \PH=(\cE mA\cd_A+\cA mI\de_I)\, \PH
                                                            \label{s14}\eeq
where $\6_m$ denote the partial derivatives, $\E am$ are the entries of the
inverse vielbein
\[ \E am\e mb=\de_a^b,\qd\e ma\E an=\de_m^n,\]
and the following notation and summation conventions are
used in order to simplify the notation:
\bea
& &\{\cE mA\}=\{\e ma,\ \p m\a,\ -\dps m\da\},\qd
\{\cA mI\}=\{\o m{ab},\ \A mi\},                             \label{s11}\\
& &\cE mA X_A=\e ma X_a+\p m\ua X_\ua,\qd
\p m\ua X_\ua=\p m\a X_\a-\dps m\da X_\da.                  \label{sc1}\eea
Requiring $[\6_m,\6_n]\PH=0$ one obtains from \Gl{s14} by means of \Gl{s5}
\bea & &\F abI=2\E am\E bn(\6_{[m}\cA {n]}I+\s0 12\f JKI\cA mJ\cA nK
            +\e {[m}c\p {n]}\ua\F \ua{c}I
            +\s0 12\p m\ua\p n\ube\F \ua\ube{I}),
                                                            \label{s16}\\
& &\T abA=2\E am\E bn(\6_{[m}\cE {n]}A+\e {[m}c\p {n]}\ua\T \ua{c}A
                        +\cE {[m}B\cA {n]}I\Gg IBA
            +\s0 12\p m\ua\p n\ube\T \ua\ube{A})
                                                            \label{s18}\eea
i.e. the curvatures $\F abI$ and the torsions $\T abA$ are given in
terms of the gauge fields, their partial derivatives and the remaining
torsions and curvatures.
The equation which determines $\T abc$ can also be solved
for $\o m{ab}$ and yields in particular:
\bea & &\T abc=0\ \LRA\
      \o m{ab}=E^{an}E^{br}(\om_{[mn]r}-\om_{[nr]m}+\om_{[rm]n}),
                                                                         \nn\\
    & &  \om_{[mn]r}=e_{ra}(\6_{[m}\e {n]}a +\e {[m}a\A {n]}\dD
                   +\e {[m}b\p {n]}\ua\T \ua{b}a
                   +\s0 12\p m\ua\p n\ube\T \ua\ube{a})
\label{s19}\eea
where $\A m\dD$ denote the gauge fields of Weyl
transformations. The BRS operator is constructed by means of
ghost fields associated with
diffeomorphisms, supersymmetry and structure group transformations and denoted
by $C^m$, $\xi^\a$, $\5\xi^\da$ and $C^I$ respectively ($C^m$, $C^I$ are
anticommuting and $\xi^\a$, $\5\xi^\da$ are commuting ghosts). The
BRS transformations read
\bea
s\,\PH      &=&(C^n\6_n+\xi^\a\cd_\a+\5\xi^\da\5\cd_\da
             +C^I\de_I)\,\PH,                               \label{s20a}\\
s\,\e ma    &=& C^n\6_n\e ma+(\6_mC^n)\e na
             +C^I\Gg Iba\e mb-\xi^\ua\cE mB\T B\ua{a},           \label{s35}\\
s\,\p m\a   &=&C^n\6_n\p m\a+(\6_mC^n)\p n\a+D_m\xi^\a
             +C^I \Gg I\be\a \p m\be
             -\xi^\ube\cE mB\T B\ube\a,                    \label{s36}\\
s\,\cA mI    &=&C^n\6_n\cA mI+(\6_mC^n)\cA nI+D_mC^I
             -\xi^\ua\cE mB\F B\ua{I},                     \label{s39}\\
s\,C^m      &=&C^n\6_nC^m+\s0 12\xi^\ug\xi^\ube\T \ube\ug{a}\E am,
                                                            \label{s40}\\
s\,\xi^\a   &=&C^n\6_n\xi^\a+C^I\Gg I\be\a\xi^\be
          +\s0 12\xi^\ug\xi^\ube(\T \ube\ug\a-\T \ube\ug{a}\E am\p m\a),
                                                              \label{s41}\\
s\, C^I      &=&C^n\6_nC^I+\s0 12 \f KJI C^JC^K
            +\s0 12\xi^\ug \xi^\ube(\F \ube\ug{I}-\T \ube\ug{a}\E am \cA mI)
                                                           \label{s44}\eea
where $D_m\xi^\a$ and $D_mC^I$ are defined by
\beann & & D_m\xi^\a=\6_m\xi^\a-\cA mI \Gg I\be\a \xi^\be,\qd
   D_mC^I=\6_mC^I+\cA mJ\f JKI C^K\eeann
and the following summation convention is used
(notice the different signs in \Gl{sc1} and \Gl{sc2})
\beq \xi^\ua X_\ua=\xi^\a X_\a+\5\xi^\da X_\da .          \label{sc2}\eeq
\Gl{s14}, \Gl{s16}, \Gl{s18} and \Gl{s20a}--\Gl{s44} can be written in
the following compact form
\bea
\4s\, \PH&=&(\4\xi^A\cd_A+\4C^I\de_I)\, \PH,                \label{s20}\\
\4s\, \4\xi^A  &=&\4C^I\Gg IBA \4\xi^B-\s0 12(-)^{|B|} \4\xi^B \4\xi^C\T CBA,
                                                          \label{s31}\\
\4s\, \4C^I  &=&\s0 12\f KJI \4C^J \4C^K-\s0 12(-)^{|A|} \4\xi^A \4\xi^B\F BAI
                                                          \label{s32}\eea
where $|A|$ denotes the grading of $\cd_A$ ($|a|=0$, $|\a|=|\da|=1$), and
$\4\xi^A$, $\4C^I$ are the generalized connection `forms' mentioned in the
introduction and defined by
\beq \4\xi^a=\7C^n\e na,\qd\4\xi^\a=\xi^\a+\7C^n\p n\a,
\qd\4\xi^\da=\5\xi^\da-\7C^n\dps n\da
         \qd \4C^I=C^I+\7C^n\cA nI                       \label{t15}\eeq
where $\7C^n$ denotes the sum of the differential $dx^n$ and the ghost
$C^n$ of diffeomorphisms:
\beq \7C^n=dx^n+C^n.                                     \label{t15a}\eeq
Using \Gl{s20}--\Gl{s32} one may verify that
\beq \4s^2=0\qd\LRA\qd s^2=[s,\6_m]=[\6_m,\6_n]=0
\label{t12}\eeq
holds by virtue of the algebra \Gl{s5} and its Bianchi identities
which have been discussed in \cite{gwz}, \cite{ggmw}, \cite{mul} and read
\bea & &\csum {ABC}{36}(-)^{|A|\, |C|}
     (\cd_A\T BCD+\T ABE\T ECD+\F ABI\Gg ICD)=0,
                                                          \label{s8a}\\
 & &    \csum {ABC}{36}(-)^{|A|\, |C|}
     (\cd_A\F BCI+\T ABD\F DCI)=0
                                                          \label{s8b}\eea
where ${\displaystyle \csum {}{15}}$ denotes the cyclic sum.

In order to construct a gauge fixed action one introduces in addition
an antighost field $\zeta$ and a `Lagrange multiplier field' $b$ for
each gauge field and defines $s\zeta=b$, $sb=0$ which implies that
these fields contribute only trivially to solutions of \Gl1.
This completes the construction of the BRS algebra
for supergravity theories whose field content consists
of tensor fields, gauge fields $\cE mA$, $\cA mI$
and the corresponding ghosts, antighosts and $b$--fields. For
supergravity theories containing additional fields
one has to specify their BRS transformations as well. This will be
done in subsection \ref{new}
for new minimal supergravity which contains antisymmetric
gauge potentials, corresponding ghosts and a ghost for these ghosts.

I stress that the BRS operator is defined and nilpotent on all fields
unlike the algebra \Gl{s5} which is
assumed to be realized only on tensor fields but not on the gauge fields
or the ghosts.

\mysection{Results}\label{res}

\subsection{Old minimal supergravity}\label{old}

Apart from the gauge, ghost, antighost and $b$ fields,
the field content of old minimal supergravity with Yang--Mills
and chiral matter multiplets consists of elementary tensor fields
\beq \{\Ps^l\}=\{M,\5M,B^a,\la_\a^i,\5\la_\da^i,D^i,
\Ph^s,\5\Ph^s,\chi_\a^s,\5\chi_\da^s,F^s,\5F^s\}\label{Ps}\eeq
where $M$,$\Ph^s$,$F^s$ are complex Lorentz scalar fields
($\5M$,$\5\Ph^s$,$\5F^s$ denote the complex
conjugated fields), $B$ is a real Lorentz vector field,
$D^i$ are real Lorentz scalar fields and $\la^i$ and $\chi^s$ are
complex spinors.
$\la^i$ and $D^i$ are the gauginos and auxiliary fields of the
super Yang--Mills multiplets of $\cg_s+\cg_a$,
$\Ph^s$,$\chi^s$,$F^s$ denote the elementary component fields of the
$s$th chiral matter multiplet, $M$ and $B$ are the auxiliary fields
of the supergravity multiplet. In the following all formulas are given
for the case in which $\cg_a$ contains $\de_\dD$ and $\de_\dR$.
The results for the cases in which
Weyl or R-transformations are not contained in the gauge
group are simply obtained by setting to zero the corresponding fields
everywhere.

Except for $\F abI$ and $\T ab\a$ which are obtained from \Gl{s16} and \Gl{s18}
and up to complex conjugation the torsions and curvatures are given by
\bea
& &
\T abc=\T \a\be{a}=\T \ua\ube\ug=
\T \ua{a}b=\F \ua\ube{i}=0,\qd \F \da{a}i=i\la^{i\,\a}\si_{a\,\a\da},\nn\\
& &\T \a\da{a}=2i\so a\a\da,\qd
\T \a{a}\da=-\s0 i8 \5M{\si_{a\,\a}}^\da, \qd
\T \a{a}\be=-i\, (\de_\a^\be B_a+B^b{\si_{ab\,\a}}^\be),\nn\\
& &\F \a\be{ab}=-\5M\osuu ab\a\be,\qd
\F \a\da{ab}=2i\, \ep^{abcd}\si_{c\,\a\da}B_d,\nn\\
& &
\F {\da}b{cd}=i\, (T^{cd\a}\si_{b\, \a\da}-2\so {[c}\a\da T^{d]}{}_b{}^\a
-2\de_b{}^{[c}\so {d]}\a\da\la^{\dD\, \a}).
                                                        \label{o1}\eea
$\o m{ab}$ is obtained from \Gl{s19}.
The BRS transformations of the ghosts and gauge fields are obtained
from \Gl{s35}--\Gl{s44}, those of the fields \Gl{Ps}
from \Gl{s20a} using table 1 where $S$, $U$, $G^{i}$, $\5S$, $\5U$, $\5G^{i}$
denote \SL2c irreducible tensors constructed of the
$\F abi$ and $\T ab\ua$:
\bea & & \G \a\be{i}=-\F abi\osuu ab\a\be,\qd
S_\a=\T ab\be\osuu ab\be\a,\qd U_{\da\dbe}{}^\g=\T ab\g
\obsuu ab\da\dbe,\qd W_{\a\be\g}=T_{ab(\a}\osuu ab\be{\g)}\nn\\
&\LRA& T_{\a\da\,\be\dbe\,\g}=\ep_{\a\be}U_{\da\dbe\g}
     +\ep_{\dbe\da}(W_{\a\be\g}+\s0 23\ep_{\g(\a}S_{\be)}),\qd
\F {\a\da\,}{\be\dbe}i=\ep_{\a\be}\5G_{\da\dbe}{}^i
   +\ep_{\da\dbe}G_{\a\be}{}^i.                           \label{s45}\eea
{\begin{center}Table 1:\end{center}
\[
\ba{c|c|c}
\Ps               &\cd_\a\Ps                   &\5\cd_\da\Ps       \\
\hline
M                 &\s0 {16}3(S_\a-i\la_\a^\dR
                            -\s0 32\la_\a^\dD) &0                  \\
B_{\be\dbe}       &\s0 13\ep_{\be\a}(\5S_\dbe+4i\5\la_\dbe^\dR)
                   -\5U_{\a\be\dbe}
                                               &\s0 13\ep_{\dbe\da}
                                                (S_\be-4i\la_\be^\dR)
                                                -U_{\da\dbe\be}    \\
\la_\be^i         &i\ep_{\a\be}D^i+\G \a\be{i} &0                  \\
D^i               &\cd_{\a\da}\5\la^{i\da}
                   +\s0 {3i}2B_{\a\da}\5\la^{i\da}
                                               &\cd_{\a\da}\la^{i\a}-
                                               \s0 {3i}2B_{\a\da}\la^{i\a}\\
\Ph              &\chi_\a                      &0                  \\
\chi_\be         &\ep_{\be\a}F                 &-2i\cd_{\be\da}\Ph \\
F                &-\s0 12\5M\chi_\a            &-2i\cd_{\a\da}\chi^\a
                                                -4\5\la_\da^i\de_i\Ph
                                                +B_{\a\da}\chi^\a
\ea\]}
Let me now describe the results of the investigation of \Gl{1}.
Up to contributions of the form $s\Ga^{-1}$ each
real local solution $\cW^0$ of \Gl1 can be written in the form\footnote{Due
to the reality of $s$ each complex solution of \Gl1 is of
the form $\cW^g=\cW^g_1+i\cW^g_2$ where $\cW^g_1$ and $\cW^g_2$ are real
solutions of \Gl1.}
\bea \cW^0_{old}=\int d^4x\, e\, \cP\, \Om+c.c.,\qd e=det(\e ma)\label{o3}
\eea
where $\cP$ is the operator
\beq \cP=\5\cd^2-4i\ps_a\si^a\5\cd-3M
+16\ps_a\si^{ab}\ps_{b}, \qd \5\cd^2=\5\cd_\da\5\cd^\da,\qd
                                       \p a\a=\E am\p m\a    \label{o4}\eeq
and $\Om$ is an antichiral function of the tensor fields given by
\beq \Om=H(\5M,\5W,\5\la,\5\Ph)+(\cd^2-\5M)\, G(\PH),\qd
\cd^2=\cd^\a\cd_\a                                           \label{o5}\eeq
where $H$ depends only on the (undifferentiated) fields $\5M,\5W_{\da\dbe\dg},
\5\la_\da^i,\5\Ph$ while $G$ depends on all variables
\beq \{\PH^r\}=\{\cd_{a_1}\ldots\cd_{a_n}\Ps^l,\
                 \cd_{a_1}\ldots\cd_{a_n}\F abI,\
                 \cd_{a_1}\ldots\cd_{a_n}\T ab\ua:\ n\geq 0\}  \label{PH}\eeq
with $\{\Ps^l\}$ as in \Gl{Ps}. $H$ and $G$ are additionally
restricted by
\bea & &\de_I H=\de_I G=0\qd\forall\,\de_I\nin\{\de_\dR,\de_\dD\},\nn\\
& &\de_\dD H=-3H,\qd \de_\dD G=-2G,\qd
   \de_\dR H=-2iH,\qd \de_\dR G=0\label{HG}\eea
where the conditions imposed by $\de_\dD$ or $\de_\dR$ of course
are absent if $\cg$ does not contain the respective generator.
I remark that $\cW^0_{old}$ contains only one
Fayet--Iliopoulos contribution \cite{fi}, namely
$ \int d^4x\, e \, a\, D^\dR$ where $a$ is a real constant.
This Fayet--Iliopoulos contribution arises from
the contribution $\0 {3a}{64}\5M$ to $\Om$ which
gives the supersymmetric version of the Einstein--Hilbert action due to
\beq \5\cd^2 \5M=\s0 {16}3(\s0 12R+2D^\dR+\s0 3{8}M\5M
-3B^aB_a+3iD^\dD-3i\cd_aB^a),\qd R=\F ab{ba}.
                                                         \label{o18}\eeq
Of course \Gl{o18} is not Weyl-invariant and thus contributes to
$\cW^0_{old}$ only in the case $\de_\dD\nin\cg$.
Using table 1 one can check that further Fayet--Iliopoulos
contributions indeed do not arise from \Gl{o3}.

Each real solution $\cW^1$ of \Gl1 is in the case of
old minimal supergravity of the form
\beq \cW^1=\cW^1_{abel}+\cW^1_{nonabel}+c.c.
\label{o6}\eeq
where $\cW^1_{abel}$ and $\cW^1_{nonabel}$ are complex solutions of
\Gl1 collecting the candidates for `abelian' and `nonabelian'
anomalies respectively. $\cW^1_{abel}$ is given by
\beq \cW^1_{abel} =\int d^4x\, e\, {\sum_j}'
\left\{C^j\cP+\xi^\a\cP^j_\a\right\}\Om_j
                                                       \label{o7}\eeq
where $\sum'_j$ runs only over the abelian factors of the gauge group,
$\cP$ is the operator \Gl{o4}, each $\Om_j$ is of the
form \Gl{o5} with $H_j$ and $G_j$ restricted by
\Gl{HG} and $\cP^j_\a$ is the operator
\beq\cP^j_\a=
4i\A aj\so a\a\da\5\cd^\da-32\A aj\osuo ab\a\be \ps_{b\be}
+8\la^j_{\a}, \qd \A aj=\E am\A mj.                 \label{o7a}\eeq
In particular \Gl{o7} contains the supersymmetric abelian
chiral anomalies which
arise from contributions $g_{kl}\5\la^k\5\la^l$ and
$a\5W_{\da\dbe\dg}\5W^{\da\dbe\dg}$
to $\Om_j$ where $g_{kl}$ are purely imaginary constant tensors of
$\cg_a+\cg_s$ and $a$ are complex constants.
Notice that if $\cg$ contains neither $\de_\dD$ nor $\de_\dR$
then $\Om_j$ contains a constant contribution $\0 {a_j}8$ which gives
rise to the following simple solutions of \Gl1:
\beq \int d^4x\, e\, {\sum_j}'a_j\, (\xi\la^j-
4\A aj\xi\si^{ab} \ps_{b}
+2C^j\ps_a\si^{ab} \ps_{b})+c.c.
                                                       \label{o7b}\eeq
$\cW^1_{nonabel}$ is a linear combination
of the independent supersymmetric nonabelian chiral anomalies
$\De^\tau_{nonabel}$
which are labeled by $\tau$ (there are as many of them as $\cg_s$ has
independent Casimir operators of third order):
\bea \cW^1_{nonabel}&=&\sum_\tau a_\tau\, \De^\tau_{nonabel},\qd
a_\tau=const.,                                         \label{nonab}\\
\De^\tau_{nonabel}&=&\int Tr\left\{ Cd(AdA+\s0 12A^3)
+(LA+AL)dA+\s0 32LA^3\right.                                 \nn\\
& &\hspace{4em}\left.
-3\, d^4x\, e\, (\5\xi\La^{\dagger}\La\La
+\xi\La\, \La^{\dagger}\La^{\dagger})\right\}               \label{o8}\eea
where $C$, $A$, $\La_\a$, $\La^{\dagger}_\da$, $L$ denote the matrices
\beq C=C^i T_i,\ A=dx^m\A mi T_i,\ \La_\a=i\la_\a^j T_j,\
\La^{\dagger}_\da=i\5\la_\da^j T_j,\ L=dx^m\e ma(\xi\si_a\La^\dagger
                                                    -\La\si_a\5\xi)
                                                       \label{o9}\eeq
defined by means of
a suitably chosen matrix representation $\{T_i\}$ of $\cg_s$ satisfying
\[ [T_i,T_j]=\f ijk T_k\]
(in general one of course has to choose different
representations $\{T_i\}$ for different values of $\tau$).
Notice that \Gl{o8}
depends on the vielbein only via the 1--forms $e^a=dx^m\e ma$ due to
\[ d^4x\, e=-\s0 1{24}\, \ep_{abcd}\, e^a e^b e^c e^d\]
i.e. the integrand of
$ \De^{\tau}$ can be written completely in terms of the forms $A^i=dx^m\A mi$,
$dA^i$, $e^a$ and the fields $C^i$, $\xi^\a$, $\5\xi^\da$, $\la^i_\a$ and
$\5\la^i_\da$. In flat space where one can choose $\e ma=\de_m^a$ and
identify $e^a$ with a constant differential $dx^a$,
\Gl{o8} agrees completely with the form of globally supersymmetric nonabelian
chiral anomalies derived in \cite{kai}, i.e. the expressions derived in
\cite{kai} are in fact not only globally but also locally supersymmetric after
replacing $dx^a$ with $e^a$, without adding any gravitino dependent terms to
these expressions. In fact the independence on the gravitino does not only hold
for $\De^\tau_{nonabel}$ but for the complete solution \Gl{d8} of the descent
equations \Gl{d3} arising from it as is shown explicitly in appendix \ref{dec}.

Notice that the
contribution $\int Tr\{ Cd(AdA+\s0 12A^3)\}$ to \Gl{o8} is the
well-known form of a nonabelian chiral anomaly in the nonsupersymmetric case
(see e.g. \cite{zu}) and
the remaining terms in \Gl{o8} represent its supersymmetric
completion.

I remark that the abelian chiral anomalies arising
from contributions $g_{kl}\5\la^k\5\la^l$ to $\Om_j$ in \Gl{o7}
(with $g_{kl}$ purely imaginary) can be written in a similar form as
\Gl{o8}. Namely
by adding a suitable counterterm $s\Ga^0$ to \Gl{o7} they become
linear combinations of solutions $\De^\tau_{abel}$ of \Gl1 given by
\bea\De^\tau_{abel}&=&\int\left\{ C^j Tr(F^2)+2A^j Tr(FL)-i\, d^4x\, e\,\left[
 \xi\la^j Tr(\La^\dagger\La^\dagger)+\5\xi\5\la^j Tr(\La\La)\right.\right.\nn\\
& &\left.\left.+2\5\la_\da^jTr(\La^{\dagger\, \da}\xi\La)
   +2\la^{j\,\a}Tr(\La_\a\5\xi\La^\dagger)\right]\right\},\qd F=dA+A^2
                                                            \label{d13}\eea
where $C^j$, $A^j$ and $\la^j$ denote the ghost, connection 1--form and
gaugino of an abelian factor of the gauge group.

The solutions $\4\om^4$ and $\4\om^5$
of the descent equations \Gl{d3} arising from \Gl{o3} and \Gl{o7} can be
written in the following remarkably simple form
\beq \4\om^4_{old}=(\4\cd_\da\4\cd^\da-M)(\Xi\, \Om),\qd
     \4\om^5_{abel}={\sum_i}'(\4\cd_\da\4\cd^\da-M)(\Xi\, \4C^i\Om_i)
                                                         \label{d4}\eeq
where $\Xi$ denotes the product of all $\4\xi^a$
and $\4\cd_\da$ is an extension of $\5\cd_\da$ suitably defined on the
variables \Gl{t15}:
\beq \Xi=-\s0 1{24}\, \ep_{abcd}\, \4\xi^a\4\xi^b\4\xi^c\4\xi^d,\qd
         \4\cd_\da\PH^r  =\5\cd_\da\PH^r,       \qd
         \4\cd_\da\4\xi^A=\4\xi^B\T B\da{A},  \qd
         \4\cd_\da\4C^i  =\4\xi^A\F A\da{i}.
                                                         \label{d6}\eeq
Using the explicit expressions \Gl{o1} one obtains in particular
\beq\4\cd_\da\4\xi_{\dbe\be}=4i\,\ep_{\dbe\da}\, \4\xi_\be,     \qd
    \4\cd_\da\4\xi_\a       =-\s0 i8\, \4\xi_{\a\da} M,         \qd
    \4\cd_\da\4C^j          =-i\,\4\xi_{\a\da}\la^{j\,\a}.
                                                         \label{d7}\eeq
The solution of \Gl{d3} arising from \Gl{o8} reads
\beq \4\om^{5,\tau}_{nonabel}=
Tr\left\{\4C\4\cF^2-\s0 12\, \4C^3\4\cF+\s0 1{10}\, \4C^5
     -3\, \Xi\, (\4\xi^\a\La_\a\,
\La^\dagger\La^\dagger+\4\xi_\da\La^{\dagger\,\da}\,
      \La\La)\right\}
                                                           \label{d8}\eeq
with $\La_\a$ and $\La^\dagger_\da$ as in \Gl{o9} and
\beq \4C=\4C^iT_i,\qd \4\cF=-\s0 12(-)^{|A|}\4\xi^A\4\xi^B\F BAi T_i.
                                                           \label{d10a}\eeq
The solution of \Gl{d3} arising from an abelian chiral
 anomaly \Gl{d13} has a similar form:
\bea \4\om^{5,\tau}_{abel,chir}=\4C^j Tr(\4\cF^2)-i\,\Xi\,\left\{
     \4\xi^\a\la_\a^j Tr(\La^\dagger\La^\dagger)+\4\xi_\da\5\la^{\da\, j}
 Tr(\La\La)\right.\nn\\
        \left.+2\5\la_\da^jTr(\La^{\dagger\, \da}\4\xi^\a\La_\a)
         +2\la^{j\,\a}Tr(\La_\a\4\xi_\da\La^{\dagger\, \da})\right\}.
                                                           \label{d12}\eea

\subsection{New minimal supergravity}\label{new}

New minimal supergravity differs from old minimal supergravity
with regard to its field content. Namely $M$,
$\la^\dR_\a$, $D^\dR$ and $B^a$ are not elementary fields
any more. Instead new minimal supergravity contains a new set of elementary
fields consisting of the
components $t_{mn}=-t_{nm}$ of a 2--form gauge potential, corresponding
ghosts $Q_m$ and a ghost $Q$ for these ghosts\footnote{In order to fix a gauge
for $t_{mn}$ and $Q_m$ one introduces additional fields whose role is analogous
to that of the fields $\zeta$ and $b$. As the latter these additional fields
contribute only trivially to solutions of \Gl1.}
which have the following reality properties, ghost numbers and gradings:
\beq\ba{lll} t_{mn}=(t_{mn})^*,& gh(t_{mn})=0,& |t_{mn}|=0,\\
             Q_m=(Q_m)^*      ,& gh(Q_m)=1   ,& |Q_m|=1,   \\
             Q=-Q^*           ,& gh(Q)=2     ,& |Q|=0.
                                                      \ea\label{t}\eeq
This field content arises from setting to zero
the field $M$ and the superfield arising from it in
old minimal supergravity with gauged R-transformations
(but without Weyl transformations).
This leads to the identifications
\bea & & M\equiv 0, \qd\la_\a^\dR\equiv -iS_\a,\qd
D^\dR\equiv -\s0 14R+\s0 32B^aB_a,                     \label{n1}\\
& &B^a\equiv\e ma\, \ep^{mnkl}\, (\s0 12\6_nt_{kl}+i\ps_n\si_k\5\ps_l)
                                                        \label{n4}\eea
where $\ep^{mnkl}=\E am\E bn\E ck\E dl\ep^{abcd}\sim 1/e$
and $\si_m=\e ma\si_a$.
\Gl{n1} and \Gl{n4} are required by $M=\cd_\a M=\cd^\a\cd_\a M=0$,
cf. table 1 and \Gl{o18}. In particular \Gl{n4} which
has been given already in the second ref. \cite{new} `solves'
$\cd_a B^a=0$ $\LRA$ $\cd^\a\cd_\a M-c.c.=0$ identically in terms of
elementary fields which is a rather involved condition
since it reads more explicitly
\beq 0=\cd_aB^a=\E am(\6_m-\s0 12\o m{bc}(e,\ps)l_{bc}-\p m\a\cd_\a
+\dps m\da\5\cd_\da)B^a            \label{n5}\eeq
where $\5\cd_\da B^a$ and $\cd_\a B^a$
are obtained from table 1 using the second identity \Gl{n1}:
\beq
\5\cd_\da B_{\be\dbe}=\ep_{\da\dbe}S_\be-U_{\da\dbe\be}\ \LRA\
    \5\cd_\da B^a=-\s0 i2\, \ep^{abcd}\si_{b\, \a\da}\T cd\a.
                                                   \label{n5a}\eeq
The nilpotent BRS transformations of $t_{mn}$, $Q_m$ and $Q$ read
\bea
s\, t_{mn} &=&C^k\6_kt_{mn}+(\6_mC^k)\, t_{kn}+(\6_nC^k)\, t_{mk}
           -\6_mQ_n +\6_nQ_m\nn\\
        & &-i(\xi\si_m\5\ps_n-\xi\si_n\5\ps_m+\ps_m\si_n\5\xi-\ps_n\si_m\5\xi),
                                                               \label{n6}\\
s\, Q_m    &=&C^n\6_nQ_m+(\6_mC^n)Q_n+\6_mQ
           +2i\,\xi\si^n\5\xi \, t_{nm}-i\,\xi\si_m\5\xi,          \label{n7}\\
s\, Q      &=&C^m\6_mQ-2i\,\xi\si^m\5\xi\, Q_m
                                                               \label{n8}\eea
where $\si^m=\E am\si^a$. \Gl{n4} and \Gl{n6}--\Gl{n8} can be written in
a more compact notation which is analogous to
\Gl{s31} and \Gl{s32} and reads
(with $\4\xi^A$ and $\7C^m$ as in \Gl{t15}, \Gl{t15a}):
\beq \4\cH=\4s\, \4Q,\qd
\4\cH=\s0 16\,\4\xi^a \4\xi^b \4\xi^c\ep_{abcd}B^d+
      i\,\4\xi^\a\4\xi_{\a\da}\4\xi^\da,\qd
     \4Q=\s0 12 \7C^m\7C^n\, t_{mn}+\7C^nQ_n+Q.
                                                              \label{n9}\eeq
The BRS transformations of the remaining fields can
be obtained from their counterparts in old minimal supergravity using
\Gl{n1} and \Gl{n4}. Therefore
one obtains solutions of \Gl1 for new minimal supergravity
simply by making the identifications \Gl{n1}, \Gl{n4} in the solutions
$\cW^0_{old}$, $\cW^1_{abel}$ and $\cW^1_{nonabel}$
obtained for old minimal supergravity.
However the presence of the fields $t_{mn}$, $Q_m$ and $Q$ is responsible
for the fact that this does not give {\it all} solutions of
\Gl1 with $g=0,1$ for new minimal supergravity. Namely there are
a few additional solutions which cannot be obtained in this way. I denote these
additional solutions by $\cW_{FI}$ in order to point out
that $\cW^0_{FI}$ are locally supersymmetric
versions of Fayet--Iliopoulos contributions to the action and $\cW^1_{FI}$
are anomaly candidates of a similar form. They are given by
\bea \cW^0_{FI}& =&\int d^4x \, e\, {\sum_i}'a_{i}
      \left\{D^i-2B^a\A ai
      +\la^i\si^m\5\ps_m+\ps_m\si^m\5\la^i\right\},
                                                              \label{n13}\\
\cW^1_{FI}&=&\int d^4x \, e\, {\sum_{i,j}}'a_{ij}
      \left\{ C^i(D^j+\ep^{mnkl}\A mj\6_nt_{kl}
      +\la^j\si^m\5\ps_m+\ps_m\si^m\5\la^j)\right.               \nn\\
  & &\left. -\A mi(\xi\si^m\5\la^j+\la^j\si^m\5\xi)
     -i\ep^{mnkl}\A mi \A nj(\xi\si_k\5\ps_l+\ps_k\si_l\5\xi)\right\},\qd
a_{ij}=-a_{ji}
                                                              \label{n15}\eea
where $\sum'_i$ and $\sum'_{i,j}$ run only over abelian factors as in \Gl{o7}
and the constants $a_i$, $a_{ij}$ must be chosen real in order
to get real solutions of \Gl1. Notice that the supersymmetrized
Einstein--Hilbert action $\cW^0_{EH}$ is contained in \Gl{n13} since $D^\dR$
which contributes to \Gl{n13} has to be identified with $-\0 14R+\0 32B^aB_a$
according to \Gl{n1}. The form of $\cW^0_{EH}$ obtained from \Gl{n13}
agrees with that given in the second ref. \cite{new} and in \cite{mul}.
Thus we obtain the result that
in the case of new minimal supergravity the action contains
a Fayet--Iliopoulos contribution for each abelian factor apart from
the one corresponding to the R-transformations which becomes a
contribution $\int d^4x e(-\0 14R+\0 32B^aB_a)$ to $\cW^0_{EH}$.

I stress again that apart from $\cW^0_{EH}$ the solutions
$\cW^g_{FI}$ do not have counterparts in old minimal
supergravity. This holds in particular for the anomaly
candidates \Gl{n15}. Notice that they are
present only if the structure group contains
at least two abelian factors, i.e. at least one abelian factor
in addition to the R-transformation, since the $a_{ij}$ are antisymmetric.

The solutions of \Gl{d3} arising from \Gl{n13} and \Gl{n15} are
\bea \4\om^{4}_{FI}&=&{\sum_i}'a_{i}\left\{2\4C^i\4\cH
+\5\la^i\5\h+\h\la^i+\Xi \, D^i\right\},
                                                              \label{d14}\\
\4\om^{5}_{FI}&=&{\sum_{i,j}}'a_{ij}\left\{\4C^i\4C^j\4\cH
+\4C^i\5\la^j\5\h+\h\la^j\4C^i+\Xi\, \4C^i D^j\right\}
                                                              \label{d15}\eea
with $\4\cH$ as in \Gl{n9} and
\beq \5\h^\da=\s0 i{6}\,\4\xi^{\da\a}\4\xi_{\a\dbe}\4\xi^{\dbe\be}\4\xi_\be,\qd
\h^\a=-\s0 i{6}\,\4\xi_\dbe\4\xi^{\dbe\be}\4\xi_{\be\da}\4\xi^{\da\a}.
                                                              \label{d15a}\eeq

\subsection{Nonminimal supergravity}\label{non}

The nonminimal supergravity theories discussed in the following are
as in \cite{ggmw} parametrized by the real number $n$ occurring in the torsions
\beq
\T \a{a}b=2n\, \de_a^b\, T_\a,\qd \T \a\be\g=(n+1)(\de^\g_{\a}\, T_{\be}
                          +\de^\g_{\be}\, T_{\a}),\qd
\T \a\da\be=(n-1)\, \de_{\a}^\be\, \5T_\da
\label{nm0}\eeq
 and are in addition characterized by the constraints
\beq \T \a\da{a}=2i\so a\a\da,\qd\T abc=\T \a\be{a}=\F \ua\ube{i}=
\F \a\be{cd}=0.
                                                    \label{s60}\eeq
A detailed discussion of \Gl{nm0} and \Gl{s60} and their
implications can be found in \cite{ggmw}.
I consider the case in which $\cg$ contains neither Weyl nor R-transformations.
Apart from the special cases $n=0$ and $n=-\0 13$ which are
excluded for simplicity
the set of elementary tensor fields is given by
\beq  T_\a, \ \SS,\ v^a,\ G^a,\ \la_\a,\ \la_\a^i,\ D^i,
                  \ \Ph^s,\ \chi_\a^s,\ F^s                \label{nPs}\eeq
and the complex conjugated fields
where $\SS$, $v^a$ and $G^a$ denote a complex scalar field, a complex and
a real vector field which in \cite{ggmw} are denoted by
$S$, $c^a+id^a$ and $b^a$.
$\la_\a$ (denoted as in \cite{ggmw}) must not be confused with
one of the gauginos $\la_\a^i$ of $\cg_s+\cg_a$ which appear as in
minimal supergravity in the curvatures
$\F \da{a}i=i\la^{i\a}\si_{a\, \a\da}$.
The field content is completed by the gauge fields, ghosts, antighosts
and the corresponding Lagrange multiplier fields where again due to
$\T abc=0$ the spin connection is not an elementary field but
determined by \Gl{s19}.
The supersymmetry transformations of the fields \Gl{nPs}
can be found in \cite{ggmw} or obtained from table 1 (by replacing there
all fields and operators by their primed counterparts, see below).
In particular one has
\beq
\cd_\a T_\be=\s0 12\ep_{\a\be}\left[\SS+(n+1)T^\g T_\g\right],\qd
\5\cd_\da T_\a=v_{\a\da},\qd
\cd_\a \SS=0,\qd  \5\cd_\da \SS=\5\la_\da.
                                                          \label{nm1}\eeq
The BRS transformations of the fields \Gl{nPs} are obtained from
\Gl{s20a}, those of the ghosts and gauge fields by \Gl{s35}--\Gl{s44}.
However it will be useful to rewrite the BRS algebra in terms of
redefined ghosts, gauge fields and supersymmetry transformations
marked by primes and defined in terms of the original (unprimed) ghosts,
gauge fields and supersymmetry transformations such that $\cd'_\a$,
$\5\cd'_\da$ and $\cd'_a$ satisfy the algebra
\[ [\cd'_A,\cd'_B\}=-T'_{AB}{}^C\cd'_C-F'_{AB}{}^I\de_I\]
where the $T'_{AB}{}^C$ and $F'_{AB}{}^I$ satisfy the con\-straints
of mi\-ni\-mal su\-per\-gra\-vi\-ty with R-trans\-for\-ma\-tions
(in fact such redefinitions
have been used also in \cite{ggmw}). This is achieved by
\beq \cd'_\a  =\cd_\a+4n\, T^\be l_{\a\be}+i\, (3n+1)\, T_\a\de_\dR,\qd
\5\cd'_\da  =\5\cd_\da+4n\, \5T^\dbe l_{\da\dbe}-i\, (3n+1)\, \5T_\da\de_\dR
                                                         \label{nm13}\eeq
where $l_{\a\be}$ and $l_{\da\dbe}=(l_{\a\be})^*$ denote the generators
of \SL2c transformations of undotted and dotted indices defined by
\bea & & l_{\a\be}=\s0 12\osuu ab\a\be l_{ab},\qd
 l_{\a\be}X_{\g}=-\ep_{\g(\a}X_{\be)},\qd l_{\a\be}X_{\dg}=0,\nn\\
& &l_{\da\dbe}=-\s0 12\obsuu ab\da\dbe l_{ab},\qd
 l_{\da\dbe}X_{\dg}=-\ep_{\dg(\da}X_{\dbe)},\qd l_{\da\dbe}X_{\g}=0
                                                     \label{sl2a}\eea
and $\de_\dR$ which originally was not contained in $\cg$
is defined according table 2
where the R-charge of the lowest component field $\Ph^s$ of the $s$th chiral
multiplet is denoted by the real number $k^s$ ($k^s$ is not fixed by
requiring \Gl{s7} unlike the R-charges of the remaining fields).
{\begin{center}Table 2:\end{center}
\[\ba{c|c|c|c|c|c|c|c|c|c|c}
\Ps & T&\SS&v&B&\la&\la^j&D^j&\Ph^s&\chi^s&F^s\\
\hline
\de_\dR\Ps&-iT&-2i \SS&0&0&i\la&i\la^j&0&-ik^s\Ph^s&-2ik^s\chi^s&-3ik^sF^s
\\
\ea\]}
The redefined covariant derivatives read
\beq  \cd'_{a}=\E {a}m(\6_m-\ps'_m{}^\a\cd'_\a+\5\ps'_m{}^\da\5\cd'_\da
                 -\s0 12\om'_m{}^{ab}l_{ab}-A'_m{}^i\de_i)
                                                            \label{nm14}\eeq
where the sum over $i$ contains the R-transformation and the redefined
gauge fields are given by
\bea e'_m{}^a&=&\e ma,
                                                                \label{nm5}\\
\ps'_m{}^\a &=& \p m\a+in\,\5T_\da\5\si_m{}^{\da\a},\qd
        \5\ps'_m{}^\da=\dps m\da-in\,\5\si_m{}^{\da\a}T_\a,
                                                                \label{nm6}\\
A'_m{}^i&=&\A mi\qd \forall\, i\neq \dR,
                                                                \label{nm7}\\
A'_m{}^\dR  &=& (3n+1)\{\s0 12\, (v+\5v)_m-\s0 12(n-1)T\si_m\5T
                   +i\5\ps_m \5T-i\ps_m T\},
                                                                \label{nm8}\\
\om'_m{}^{ab}&=&
                E^{an}E^{br}(\om'_{[mn]r}-\om'_{[nr]m}+\om'_{[rm]n}),
                                                                         \nn\\
\om'_{[mn]r}&=&e_{ra}\6_{[m}\e {n]}a
                   -i\ps'_m\si_r\5\ps'_n+i\ps'_n\si_r\5\ps'_m
                                                            \label{nm9}\eea
with $v_m=\e ma v_a$. One can check that $T'_{AB}{}^C$ and $F'_{AB}{}^I$ indeed
are given by \Gl{o1} where of course now unprimed fields have to
be replaced by primed fields. In particular one finds
\bea          M' &=& -4n\left[\5\SS+2(3n+1)\5T\5T\right],
                                                                \label{nm16}\\
\la'{}_\a^{\, i} &=&   \la_\a^i\qd \forall\, i\neq\dR,
                                                                \label{nm16a}\\
\la'{}_\a^{\, \dR} &=& -i\, S'_\a+3ni \left[-\s0 14\,\la_\a
                    +(3n+1)\5v_{\a\da}\5T^\da+\s0 {3n+1}2\5\SS T_\a
                    +(3n+1)^2T_\a\5T\5T \right]
                                                              \label{nm17}\eea
where $S'_\a$ is obtained from \Gl{s18} using primed gauge fields and torsions:
\[ S'_\a=T'_{ab}{}^\be(e,\ps',\om',A'^\dR,M',B')\, \osuu ab\be\a.\]
Now we choose redefined ghosts such that
\beq  s=C^m\6_m+\xi^\a\cd_\a+\5\xi^\da\5\cd_\da+C^I\de_I
      = C'{\,}^m\6_m+\xi'{\,}^\a\cd'_\a+\5\xi'{\,}^\da\5\cd'_\da+C'{\,}^I\de_I
                                                            \label{nm4}\eeq
holds identically in the fields where $C'{}^I\de_I$
contains $\de_\dR$ unlike $C^I\de_I$, i.e. in particular
all terms containing $\de_\dR$ must cancel on the r.h.s. of \Gl{nm4}.
One easily verifies that this leads to
\bea & & C'{\,}^m=C^m,\qd \xi'{\,}^\a=\xi^\a,\qd \5\xi'{\,}^\da=\5\xi^\da,\qd
         C'{\,}^i=C^i\qd \forall\, i\neq \dR,
                                                              \label{nm10}\\
     & & C'{\,}^\dR=-i\, (3n+1)(\xi T+\5\xi\5T),
                                                                \label{nm11}\\
     & & C'{\,}^{ab}=C^{ab}+4n(\xi\si^{ab}T-\5\xi\5\si^{ab}\5T).
                                                              \label{nm12}\eea
Notice that $C'{\,}^\dR$ does not have a part which is linear in the
elementary fields as a consequence of the fact that $\cg$ does
not contain $\de_\dR$. I remark that analogously to \Gl{nm4}
\bea & &\6_m=\cA mA \cd_A+\cA mI \de_I=
                \ca'{}_m{}^A \cd'_A+\ca'{}_m{}^I \de_I,
                                                            \label{nm3}\\
& &\4s=\4\xi^A\cd_A+\4C^I\de_I=\4\xi'{\,}^A\cd'_A+\4C'{\,}^I\de_I,
                                                            \label{nm3a}\\
& &\4\xi'{\,}^a=\7C^n\e na,\qd \4\xi'{\,}^\ua=\xi'{\,}^\ua+\7C^n\ps'_n{}^\ua,
\qd \4C'{\,}^I=C'{\,}^I+\7C^n\ca'_n{}^I                  \label{nm4a}\eea
hold identically in the fields (all contributions containing
$\de_\dR$ cancel on the r.h.s. of \Gl{nm3} and \Gl{nm3a}) with $\7C^n=C^n
+dx^n$. $\4s^2=0$ implies that \Gl{s35}--\Gl{s44} hold for
the primed ghosts and gauge fields (with primed torsions and curvatures),
including $C'{}^\dR$ and $A'_m{}^\dR$ as one can verify explicitly:
\bea s\, A'{}_m{}^\dR&=&C^n\6_nA'{}_m{}^\dR+(\6_mC^n)A'{}_n{}^\dR
                  +\6_mC'{}^\dR+i(\la'{}^\dR\si_m\5\xi-\xi\si_m\5\la'{}^\dR),
                                                            \label{sar}\\
     s\, C'{}^\dR&=&C^n\6_nC'{}^\dR-2i\xi\si^m\5\xi \, A'{}_m{}^\dR.
                                                            \label{scr}\eea
Thus on primed fields
the BRS operator has the same form as in minimal supergravity and therefore
one obtains solutions of $s\cW^g=0$ for nonminimal supergravity from
those obtained for old minimal supergravity by replacing in the expressions
for the latter all operators and fields with primed ones.
For instance one obtains from \Gl{o4} invariant contributions to
the action for nonminimal supergravity given by
\beq  \int d^4x\, e\, \cP'\, \Om'+c.c.,\qd
\Om'=H(\5W',\5\la',\5M',\5\Ph)+(\cd'{}^2-\5M')G(\PH')
                                                              \label{nm20}\eeq
where $\cP'$ is obtained from \Gl{o4}:
\bea \cP'&=&\cP(\5\cd',\ps',M')
=\5\cd'{}^2-4i\ps'_a\si^a\5\cd'-3M'+16\ps'_a\si^{ab}\ps'_{b}\label{nm21}\\
&=&\5\cd^2-4i\ps_a\si^a\5\cd+16\ps_a\si^{ab}\ps_b +(13n-3)\5T\5\cd
               +2(3n-1)(\5\SS+8n\5T\5T-4i\ps_a\si^a\5T).
                                                              \nn\eea
I stress that $H$ and $G$ have to satisfy \Gl{HG} where of course
the condition imposed by $\de_\dD$ is absent unlike the condition
imposed by $\de_\dR$ which must hold
(otherwise \Gl{nm20} is not invariant except for the
case $n=-\0 13$).
Contributions to $H$ are therefore e.g.
\beq H(\5W',\5\la',\5M',\5\Ph)=a \5M'
+b\, \5W'{}_{\da\dbe\dg}\5W'^{\da\dbe\dg}+g_{ij}\5\la'^i\5\la'^j
+V(\5\Ph)+\ldots
                                                              \label{nm23}\eeq
where $a$ and $b$ are constants, $g_{ij}$ are invariant symmetric
tensors of $\cg$ and the superpotential $V(\5\Ph)$ has to satisfy
\beq V(\5\Ph)=\sum_{r\geq 1}
\sum_{s_1\ldots s_r}d_{s_1\ldots s_r}\5\Ph^{s_1}\ldots\5\Ph^{s_r},
\qd \de_IV(\5\Ph)=0\qd\forall\, \de_I\neq\de_\dR,\qd
\de_\dR V(\5\Ph)=-2iV(\5\Ph).\label{nm24}
\eeq
Notice that this requires that a monomial
$\5\Ph^{s_1}\ldots\5\Ph^{s_r}$ contributing to $V$ satisfies
\beq \sum_{s=s_1}^{s_r}k^s=-2
                                                             \label{nm24a}\eeq
which cannot be fulfilled if one imposes e.g. the
chirality condition $\5\cd_\da\Ph^s=0$ on each of the chiral multiplets.
Thus one has to admit non zero
$k^s$ in order to allow for solutions of \Gl{nm24a}. This
corresponds to a relaxed chirality condition
imposed on the chiral multiplets:
\beq \5\cd'_\da\, \Ph^s=0\qd \LRA\qd \5\cd_\da\, \Ph^s=(3n+1)\, k^s\,\5T_\da\,
\Ph^s.
                                                              \label{nm18}\eeq
Replacing unprimed fields and operators by primed ones
in \Gl{o7}, \Gl{o8} and \Gl{d13} one obtains solutions of $s\cW^1=0$ for
nonminimal supergravity. Of course it is not guaranteed that
these solutions are nontrivial since nonminimal
supergravity has an extended field content and thus allows for counterterms
which do not exist in minimal supergravity. The
nonabelian chiral anomalies stay nontrivial. Notice that they keep
their explicit form \Gl{o8} because
all fields contributing to \Gl{o8} agree with their primed counterparts.
Among the solutions of $s\cW^1=0$ obtained from \Gl{o7} there are in particular
those arising from $\int d^4x e\, (C^\dR\cP+\xi^\a\cP_\a^\dR)\Om_\dR$.
They are special since they do not depend on the ghosts of the structure
group due to \Gl{nm11}. Most of these solutions turn out to be
trivial but there are also nontrivial ones among them. For instance the
following (complex) solutions can be shown to be nontrivial:
\beq \cW^1_{susy}=
\int d^4x\, e\, \{(3n+1)(\xi T+\5\xi\5T)\cP'
+i\xi^\a\cP'{}_\a^\dR\}\{g_{jk}\5\la^j\5\la^k+V(\5\Ph)\}
\label{nm25}\eeq
with $V$ as in \Gl{nm24}, $\cP'$ as in \Gl{nm21} and
\beq \cP'{}_\a^\dR=
4iA'_a{}^\dR\so a\a\da\5\cd'{}^\da-32A'_a{}^\dR\osuo ab\a\be \ps'_{b\be}
+8\la'{}^\dR_{\a}.\eeq
The solution of \Gl{d3} arising from \Gl{nm25} is obtained from \Gl{d4} and
reads
\beq \4\om^5_{susy}=(\4\cd'{}^2-M')\left(\Xi\, \4C'{}^\dR
                    \{g_{jk}\5\la^j\5\la^k+V(\5\Ph)\}\right).\label{nm31}\eeq

\mysection{Verification of $\4s\, \4\om^{g+4}=0$}\label{ver}

In order to verify that \Gl{d4}, \Gl{d8}, \Gl{d12}, \Gl{d14}, \Gl{d15}
and \Gl{nm31} indeed solve \Gl{d3} it is more convenient to use a decomposition
of $\4\om^{g+4}$ into parts $\7\om_p$ of definite degree in the $\4\xi^a$
than the decomposition \Gl{d3a}:
\beq \4\om^{g+4}=\sum_{p=p_0}^4\7\om_p,
     \qd \7\om_p=\s0 1{p!}\,\4\xi^{a_1}\ldots\4\xi^{a_p}
                 \om_{a_1\ldots a_p}(\4\xi^\a,\4\xi^\da,\4C^I,\PH^r).
                                                              \label{d18}\eeq
As an example the decomposition of \Gl{d8} is given in appendix \ref{dec}.
The decomposition \Gl{d18}
starts at a degree $p_0$ which in general is different from zero
(unlike the decomposition \Gl{d3a} which always starts at degree 0) and
the part $\7\om_{p_0}$ solves
\beq \de_-\7\om_{p_0}=0                                  \label{d24}\eeq
where $\de_-$ is the operator
\beq \de_-=4i\, \4\xi^\a \4\xi^\da\0 \6{\6 \4\xi^{\da\a}}
=2i\, \4\xi^\a\so a\a\da \4\xi^\da\0 \6{\6 \4\xi^a}.
                                                          \label{d25}\eeq
\Gl{d24} follows from the fact that $\de_-$ is the only part of $\4s$
which decreases the degree in the $\4\xi^a$ (notice that
this is due to the constraints $\T \a\be{a}=0$ and $\T \a\da{a}=2i\so a\a\da$).
The knowledge of the cohomology of $\de_-$ will be very useful in the
following. It has been computed in the first ref. \cite{dok} (this derivation
is given in appendix \ref{de}) and recently in \cite{dix}. The result is
\msk

{\it Cohomology of $\de_-$:}
\beq \de_-f(\4\xi^A)=0\ \LRA\ f(\4\xi^A)=P(\5\dh^\da,\4\xi^\a)+
Q(\dh^\a,\4\xi^\da)+k\, \TH+\de_-g(\4\xi^A)
                                                          \label{d26}\eeq
where $k$ does not depend on the $\4\xi^A$, and $\dh^\a$, $\5\dh^\da$ and
$\TH$ are given by
\beq \dh^\a=\4\xi_\da\4\xi^{\da\a},\qd \5\dh^\da=\4\xi^{\da\a}\4\xi_\a,
     \qd \TH=\4\xi_\da\4\xi^{\da\a}\4\xi_\a.
                                                           \label{d27}\eeq
Notice that the $\dh$'s anticommute. Therefore the
$\de_-$-cohomology does not contain nontrivial elements with
larger degree than 2 in the $\4\xi^a$. This implies the following two
useful lemmas:\msk

{\it Lemma 1:} Each $\4s$-invariant 4--form in the $\4\xi^a$ vanishes:
\beq \4s\,\7\om_4=0\qd \LRA\qd \7\om_4=0 .                  \label{d28}\eeq

{\it Proof:} $\4s\7\om_4=0$ implies $\de_-\7\om_4=0$, cf. \Gl{d24}.
This implies $\7\om_4=\4\de_-\7\h$ for some $\7\h$ according to \Gl{d26}.
However this requires that $\7\h$ has degree 5 in the $\4\xi^a$ and thus
vanishes.\msk

{\it Lemma 2:} Each solution of \Gl{d3} whose decomposition
\Gl{d18} reads $\4\om=\7\om_3+\7\om_4$ can be written
as $\4s\7\h_4$ where $\7\h_4$ is a 4--form in the $\4\xi^a$.
$\7\h_4$ is unique and the solution of $\7\de_-\7\h_4= \7\om_3$.
\beq \4s\, (\7\om_3+\7\om_4)=0\qd\LRA\qd\exists\, \7\h_4:\qd
      \4s\,\7\h_4=\7\om_3+\7\om_4,\qd \de_-\,\7\h_4=\7\om_3.
                                                           \label{d29}\eeq

{\it Proof:} (i) Existence of $\7\h_4$: $\4s(\7\om_3+\7\om_4)=0$ implies
 $\de_-\7\om_3=0$ according to \Gl{d24}. By means of \Gl{d26} one concludes
the existence of $\7\h_4$
such that $\de_-\7\h_4=\7\om_3$. $\4s\7\h_4=\7\om_3+\7\om_4$ follows from
the fact that $\4\om':=\7\om_3+\7\om_4
-\4s\7\h_4$ is a $\4s$-invariant 4--form
in the $\4\xi^a$ and thus vanishes according to lemma 1.
(ii) Uniqueness of $\7\h_4$: Assume that
$\7\h_4$ and $\7\h'_4$ are two 4--forms in the $\4\xi^a$
whose $\4s$ transformations equal
 $\7\om_3+\7\om_4$. Then the difference $\7\h_4-\7\h'_4$ is a $\4s$-invariant
4--form and thus vanishes according to lemma 1.

I use these lemmas in the following in order to prove that
$\4\om^5_{abel}$, $\4\om^{5,\tau}_{nonabel}$ and $\4\om^4_{FI}$
solve \Gl{d3}. The $\4s$-invariance of $\4\om^4_{old}$,
 $\4\om^{5,\tau}_{abel,chir}$ and $\4\om^5_{FI}$ can be proved
completely analogously, the $\4s$-invariance of \Gl{nm31} is a special
case of $\4s\4\om^5_{abel}=0$ after replacing unprimed quantities by
their primed counterparts.

\subsection{Proof of $\4s\,\4\om^5_{abel}=0$}

I first note that the decomposition \Gl{d18} of $\4\om^5_{abel}$ reads
\beq \4\om^5_{abel}=\7\om_2+\7\om_3+\7\om_4 ,\qd
      \7\om_2=-4i{\sum_j}'\5\dh_\da\5\dh^\da \4C^j\, \Om_j
               \label{d30}\eeq
where $\7\om_2$ has been evaluated by means of \Gl{de6}.
Furthermore one easily verifies by means of \Gl{d7}
that $\4\om^5_{abel}$ does not depend on $\4\xi^\da$:
\beq \4\om^5_{abel}=\4\om^5_{abel}(\4\xi^a,\4\xi^\a,\4C^i,\PH^r).
                                                          \label{d33}\eeq
Using \Gl{s20}--\Gl{s32} one can check that
\beq \mbox{on $\PH^r$, $\4\xi^A$ and abelian $\4C^i$:}\qd
 \4s=\4\xi^A\4\cd_A+\4C^I\de_I,\qd \de_I\4\xi^A=\Gg IBA\4\xi^B
                                                         \label{d34}\eeq
with $\4\cd_\da$ given by \Gl{d7} and $\4\cd_a$ and $\4\cd_\a$ defined by
\beq \ba{llll}\4\cd_a\PH^r=\cd_a\PH^r,& \4\cd_a\4\xi^b=0,&
              \4\cd_a\4\xi^\ua=\s0 12\4\xi^b\T ab\ua+\4\xi^\be\T a\be\ua,&
              \4\cd_a\4C^i=\s0 12\4\xi^b\F abi+\4\xi^\a\F a\a{i},\\
              \4\cd_\a\PH^r=\cd_\a\PH^r,& \4\cd_\a\4\xi^a=0,&
              \4\cd_\a\4\xi^\ua=0,&
              \4\cd_\a\4C^i=0.\ea                         \label{d35}\eeq
An important property of the operators $\4\cd_\da$ is that they
can be used to construct a `chiral projector' $(\4\cd_\da\4\cd^\da-M)$
for functions which do not carry dotted spinor indices and do not
depend on $\4\xi^\da$:
\beq l_{\da\dbe}\, f(\4\xi^a,\4\xi^\a,\4C^i,\PH^r)=0\ \then\
\4\cd_\da(\4\cd_\dbe\4\cd^\dbe-M)f(\4\xi^a,\4\xi^\a,\4C^i,\PH^r)=0
                                                           \label{d36}\eeq
where $l_{\da\dbe}=(l_{\a\be})^*$ generates
the \SL2c transformations of dotted spinor indices, cf. \Gl{sl2a}.
\Gl{d36} follows from the fact that the $\4\cd_\da$ satisfy on the
variables $\PH^r$, $\4\xi^a$, $\4\xi^\a$ and $\4C^i$ the same
algebra as $\5\cd_\da$ on the $\PH^r$, namely $\{\4\cd_\da,\4\cd_\dbe\}
=-M l_{\da\dbe}$ as one easily checks by means of \Gl{d7}.
 Using $\de_I\4\om^5_{abel}=0$ one concludes by
means of \Gl{d34} and \Gl{d36} that $\4s\4\om^5_{abel}$ is given by
\beq \4X:=\4s\, \4\om^5_{abel}=(\4\xi^a\4\cd_a+\4\xi^\a\4\cd_\a)\,
                               \4\om^5_{abel}.              \label{d38}\eeq
Now I prove that each term occurring in the decomposition
\Gl{d18} of $\4X$ vanishes. \Gl{d30}, \Gl{d35} and \Gl{d38} imply that
this decomposition reads $\4X=\7X_2+\7X_3+\7X_4$. First one verifies
\beq \7X_2=\4\xi^\a\4\cd_\a\7\om_2
          =4i\,\5\dh_\da\5\dh^\da{\sum_j}'\4C^j\, \4\xi^\a\cd_\a\Om_j=0
                                                           \label{d39}\eeq
which holds due to $\cd_\a\Om_j=0$ and implies
$\4X=\7X_3+\7X_4$. Since $\4X$ is $\4s$-invariant by construction
one concludes by means of lemma 2 that there is a 4--form $\7\h_4$ such that
$\de_-\7\h_4=\7X_3$.
On the other hand one easily verifies that $\4X$ does not depend on
$\4\xi^\da$ since this holds already for $\4\om^5_{abel}$, cf. \Gl{d33},
and since none of the $\4\cd_A$ transformations \Gl{d35}
depends on $\4\xi^\da$. This however contradicts $\de_-\7\h_4=\7X_3$
unless $\7X_3=0$ since $\de_-\7\h_4$
of course depends both on $\4\xi^\a$ and $\4\xi^\da$
unless it vanishes. One concludes $\4X=\7X_4=0$ by means of lemma 1
which proves $\4s \4\om^5_{abel}=0$.

\subsection{Proof of $\4s\,\4\om^{5,\tau}_{nonabel}=0$}

In order to prove that \Gl{o8} solves \Gl{d3} I show that
\bea & & \4s\,\4q=Tr(\4\cF^3),   \qd
         \4q=Tr(\4C\4\cF^2-\s0 12\, \4C^3\4\cF+\s0 1{10}\, \4C^5),
                                                           \label{d9a}\\
     & & \4s\,\4p=Tr(\4\cF^3),\qd \4p=3\, \Xi\, Tr(\4\xi^\a\La_\a\,
             \La^\dagger\La^\dagger+\4\xi_\da\La^{\dagger\,\da}\, \La\La)
                                                           \label{d9b}\eea
which implies $\4s\,\4\om^{5,\tau}_{nonabel}=0$ due to
$\4\om^{5,\tau}_{nonabel}=
\4q-\4p$. \Gl{d9a} is easily verified by means of
\beq \4s\, \4C=-\4C^2+\4\cF,\qd \4s\, \4\cF=\4\cF \4C-\4C\4\cF
                                                          \label{d11}\eeq
which follows immediately from \Gl{s32} and $\4s^2=0$.
\Gl{d9b} is verified using
\beq \4\cF=\s0 12\, \4\xi^a\4\xi^b\F abi T_i-\dh^\a\La_\a
            -\5\dh_\da\La^{\dagger\, \da}
                                                              \label{d16}\eeq
which is obtained by inserting the explicit expressions for
$\F ABi$ given in \Gl{o1} into \Gl{d10a}. Using
$\dh^\a \dh^\be \dh^\g$=$\5\dh^\da \5\dh^\dbe \5\dh^\dg$=0
one easily verifies that the decomposition \Gl{d18} of $Tr(\4\cF^3)$ reads
\beq Tr(\4\cF^3)=\7\om_3+\7\om_4,\qd
\7\om_3=\s0 32\left\{\dh^\a\dh_\a\5\dh_\da
     Tr(\La^{\dagger\,\da}\,\La\La)
     +\5\dh_\da\5\dh^\da\dh^\a Tr(\La_\a\, \La^\dagger\La^\dagger)\right\}.
                                                          \label{d42}\eeq
(Notice that $\4\cF$ does not contain a part
of degree 0 in the $\4\xi^a$ due to the contraints $\F \ua\ube{i}=0$).
Due to $\4s\, Tr(\4\cF^3)=0$ one concludes by means of lemma 2 that in order
to prove \Gl{d9b} one only has to verify $\de_-\4p=\7\om_3$ which holds
due to \Gl{de4}.

\subsection{Proof of $\4s\,\4\om^4_{FI}=0$}

In order to calculate $\4s\,\4\om^4_{FI}$ one uses
\beann & &s\, \4C^j=\4\cF^j=\s0 12\, \4\xi^a\4\xi^b\F abj-i\dh^\a\la_\a^j
            -i\5\dh_\da\5\la^{j\, \da}\nn\\
& \then & \4s\,(\4C^j\4\cH)=\4\cF^j\4\cH=i\4\cF^j\TH+(\mbox{4--form in
          $\4\xi^a$})                                          \nn\eeann
which holds for abelian $\4C^j$ due to \Gl{s32} and \Gl{n9}.
Using in addition \Gl{de3} one easily checks
that the decomposition \Gl{d18} of $\4s\,\4\om^4_{FI}$ is given by
\beann & &\4Y:=\4s\,\4\om^4_{FI}={\sum_j}'a_j\, (\7Y^j_2+\7Y^j_3)+\7Y_4,\nn\\
& &\7Y^j_2+\7Y^j_3=2i\4\cF^j\TH-2\5\la^j\5\dh\TH-2\dh\la^j\TH+
\4\xi^\ua(\cd_\ua\5\la_\dbe^j)\5\h^\dbe-\h^\be\4\xi^\ua(\cd_\ua\5\la_\be^j)
-\s0 13\,  \dh^\a\4\xi_{\a\da}\5\dh^\da D^j.
                                                              \eeann
By means of the explicit form for $\4\cF^j$ given above and the $\cd_\ua$
transformations
of $\la^j$ given in table 1 one verifies $\7Y^j_2+\7Y^j_3=0$
which implies $\4Y=\7Y_4$. By means of lemma 1 one finally concludes
$\4Y=\4s\,\4\om^4_{FI}=0$.

\mysection{Conclusion}

The result of the investigation of \Gl1 for $g=0$ contains
the statement that in the case of old minimal supergravity the action
contains at the most one Fayet--Iliopoulos contribution,
namely the one which corresponds to the R-transformations, while
in the case of new minimal supergravity the action contains a Fayet--Iliopoulos
contribution for each abelian factor of the gauge group apart from the one
which corresponds to the R-transformations.

Among the solutions of \Gl1 with
$g=1$ there are in the case of new minimal supergravity
up-to-now unknown anomaly candidates \Gl{n15}
which do not have counterparts in old minimal supergravity and are present
if the gauge group contains at least two abelian factors.
Furthermore the result contains the statement that local supersymmetry
itself is not anomalous in the cases of old and new minimal supergravity
since there are no solutions of \Gl1 for $g=1$ in these cases which depend
only on the supersymmetry ghosts and the ``classical fields'' but not
on the ghosts $C^I$ of the structure group. I remark that this result follows
from the QDS structure \cite{glusy}, \cite{dok} of the theories
discussed in subsections \ref{old} and \ref{new} and may become invalid
for instance if one drops the assumption that all matter multiplets are
chiral multiplets and allows for matter
multiplets which do not have QDS structure.

Of course not only the matter multiplets must have QDS structure in order
to guarantee that supersymmetry itself is not anomalous but this must hold
also for the supersymmetry multiplets containing the torsions $\T ABC$ and
the curvatures $\F ABI$.
Examples for supergravity theories where these multiplets do not have
QDS-structure are the nonminimal theories discussed in subsection \ref{non}.
They indeed allow for solutions of \Gl1 which do not depend on the $C^I$ and
which therefore provide candidates for anomalies of local
supersymmetry. Some of these solutions have been given explicitly in \Gl{nm25}.
They seem to have close relationships to candidates for anomalies of
R-transformations present in minimal supergravity as their construction given
in subsection \ref{non} suggests.

\appendix

\mysection{Cohomology of $\de_-$}\label{de}

In order to prove \Gl{d26} we
first determine those $\de_-$-invariant functions which do not depend
on $\4\xi^\a$ (notice that they are nontrivial since each nonvanishing
function of the form $\de_-Y$ depends both on $\4\xi^\a$ and $\4\xi^\da$).
The result is:
\beq\de_{-}f(\4\xi^{\da\a},\4\xi^\da)=0\qd\LRA\qd
f=Q(\dh^\a,\4\xi^\da).\label{te4}\eeq
In order to prove \Gl{te4} we choose a basis for the functions
$f(\4\xi^{\da\a},\4\xi^\da)$ consisting of the following mutually distinct
subsets
$f_{i,n}$, $i=1,\ldots,10$, $n=0,1,\ldots$:
\beann & &f_{1,n}=\{\4\xi^{\da_1}\ldots\4\xi^{\da_n}\},\\
& &f_{2,n}=\{\4\xi^{(\da_1}\ldots\4\xi^{\da_n}\4\xi^{\dbe)\be}\},\qd
f_{3,n}=\{\4\xi^{\da_1}\ldots\4\xi^{\da_{n}}\4\xi_\dbe \4\xi^{\dbe\be}\},\\
& &f_{4,n}=\{\4\xi^{\da_1}\ldots\4\xi^{\da_n}{\4\xi^\a}{}_\dbe
\4\xi^{\dbe\be}\},\qd
f_{5,n}=\{\4\xi^{(\da_1}\ldots\4\xi^{\da_n}\4\xi^{\dbe}{}_\a\4\xi^{\dg)\a}\},\\
& &f_{6,n}=\{\4\xi^{(\da_1}\ldots\4\xi^{\da_{n}}\4\xi^{\dbe)}{}_\a\4\xi_\dg
\4\xi^{\dg\a}\},\qd f_{7,n}=\{
\4\xi^{\da_1}\ldots\4\xi^{\da_{n}}\4\xi_\dg
\4\xi^{\dg\a}\4\xi_\dbe{\4\xi_\a}{}^{\dbe}\},\\
&
&f_{8,n}=\{\4\xi^{(\da_1}\ldots\4\xi^{\da_n}\4\xi^{\dbe)\g}\4\xi_{\g\dg}\4\xi^{\dg\a}\},\qd
f_{9,n}=\{\4\xi^{\da_1}\ldots\4\xi^{\da_{n}}\4\xi_\dbe
\4\xi^{\dbe\g}\4\xi_{\g\dg}\4\xi^{\dg\a}\},\\
& &f_{10,n}=\{\4\xi^{\da_1}\ldots\4\xi^{\da_n}
\4\xi^{\dbe\be}\4\xi^{\dg\g}\4\xi_{\be\dg}\4\xi_{\g\dbe}\}
                                                                  \eeann
where each $f_{i,n}$ denotes a complete irreducible \SL2c multiplet,
e.g. $f_{1,1}=\{\4\xi^{\dot 1},\4\xi^{\dot 2}\}$. Since $\de_-$ commutes
with the \SL2c generators and changes the degree in the $\4\xi^{\da\a}$ resp.
$\4\xi^\da$ by 1 resp. $-1$, one can investigate without loss of
generality each subspace $f_{i,n}$ separately in order to determine
the $\de_-$-invariant functions $f(\4\xi^{\da\a},\4\xi^\da)$.
One easily verifies that only the functions contained in
$f_{1,n}$, $f_{3,n}$ and $f_{7,n}$ are $\de_-$-invariant and that
they are those which are of the form $Q(\dh^\a,\4\xi^\da)$. This proves
\Gl{te4}. Analogously one can show that the $\de_-$-invariant functions
which do not depend on $\4\xi^\da$ are of the form $P(\5\dh^\da,\4\xi^\a)$.

Now we investigate those functions $f(\4\xi^A)$ which vanish
for $\4\xi^\a=0$ and for $\4\xi^\da=0$:
\beq f=\sum_{q,r,s,t}a_{qrst}(\4\xi^{\da\a})\, (\4\xi^1)^q
(\4\xi^2)^r(\4\xi^{\dot 1})^s(\4\xi^{\dot 2})^t,\qd a_{00st}=a_{qr00}=0.
                                                        \label{t13}\eeq
Without loss of generality one can assume that $f(\4\xi^A)$ has definite
ghost number and thus can be written as a polynomial
in $\4\xi^1$ with coefficients which depend on
$\4\xi^2$, $\4\xi^{\da\a}$ and $\4\xi^\da$:
\beq
f(\4\xi^A)=\sum_{n=0}^{\on}(\4\xi^1)^nf_n(\4\xi^2,\4\xi^{\da\a},\4\xi^\da).
                                                          \label{t13a}\eeq
$\de_-$ is decomposed into
\beq \de_-=\4\xi^\a \4D_\a,\qd \4D_\a=4i\,\4\xi^\da\0 \6{\6\4\xi^{\da\a}}.
                                                                     \eeq
$\de_{-}f=0$ implies in particular
\beq \4D_1f_\on=0.                                        \label{t13b}\eeq
In order to solve \Gl{t13b} we define an operator $r$ whose
anticommutator with $\4D_1$ is the counting operator $\cN$ for the
variables $\4\xi^{\da 1}$ and $\4\xi^\da$, $\da=\dot 1,\dot 2$:
\beq r=-\s0 i4\, \4\xi^{\da 1}\0 \6{\6\4\xi^\da}\ \then\
\{r,\4D_1\}=\4\xi^{\da 1}\0 \6{\6\4\xi^{\da 1}}+\4\xi^\da\0 \6{\6\4\xi^\da}
=:\cN.                                              \label{t13c}\eeq
Due to \Gl{t13} $f$ does not contain a zero mode of $\cN$. Therefore one
concludes by means of standard arguments (for instance by means of the
Basic lemma \cite{let}) that $f_\on$ is a trivial solution of \Gl{t13b}
(notice that $(\4D_1)^2=0$):
\beq f_\on=\4D_1g_{\on}(\4\xi^2,\4\xi^{\da\a},\4\xi^\da).      \label{t13d}\eeq
This implies due to $\4\xi^1\4D_1=\de_{-}-\4\xi^2\4D_2$:
\beq \on\neq 0:\qd f=\de_{-}g+f',\qd f'=\sum_{n=0}^{\on-1}(\4\xi^1)^nf'_n
(\4\xi^2,\4\xi^{\da\a},\4\xi^\da)
                                                       \label{t13e}\eeq
where
\beann g=(\4\xi^1)^{\on-1}g_\on,\qd
f'_{\on-1}=f_{\on-1}-\4\xi^2\4D_2g_\on,\qd n<\on-1:\qd f'_n=f_n. \eeann
Notice that $f'$ is a polynomial which has lower degree in $\4\xi^1$
than $f$ and differs from $f$ only by a trivial contribution. Therefore
one can iterate the argument leading to \Gl{t13e} and conclude that
a $\de_-$-invariant polynomial of $\4\xi^1$ is trivial up to a part $F$ which
does not depend on $\4\xi^1$ at all. $F$ is written as a polynomial
in $\4\xi^2$ with coefficients $F_r(\4\xi^{\da\a},\4\xi^\da)$ ($F_0$ vanishes
due to \Gl{t13}):
\beq f=\de_{-}h+F,\qd F=\sum_{r\geq 1}(\4\xi^2)^rF_r(\4\xi^{\da\a},\4\xi^\da).
                                                        \label{t13f}\eeq
$\de_-f=0$ requires
\beq \de_- F_r(\4\xi^{\da\a},\4\xi^\da)=0\qd \forall r      \label{t13g}\eeq
since $\de_-$-invariant functions of different degree in the $\4\xi^\a$
have to be separately invariant (thus in fact one can assume without
loss of generality that the sum $\sum_r$ in \Gl{t13f} contains only one
nonvanishing contribution). By means of \Gl{te4} one concludes from
\Gl{t13g}
\beq F_r(\4\xi^{\da\a},\4\xi^\da)=Q_r(\dh^\a,\4\xi^\da).       \label{t13h}\eeq
Thus $F$ is a linear combination of $\de_-$-invariant monomials
$(\4\xi^2)^r$$(\4\xi^{\dot 1})^s$$(\4\xi^{\dot 2})^t$$(\dh^1)^v$$(\dh^2)^w$
where $v,w\in\{0,1\}$ since the $\dh^\a$ anticommute.
By means of the identities
\bea & &\4\xi^2\4\xi^\da=-\s0 i4\, \de_{-}\4\xi^{\da 2} ,\qd
\4\xi^2\dh^2=\4\xi^2 \4\xi^{\da 2}\4\xi_\da=
-\s0 i8\, \de_{-}(\4\xi^2{}_\da \4\xi^{\da 2}),\label{A11a}\\
& &(\4\xi^2)^2\dh^1=(\4\xi^2)^2\4\xi^{\da 1}\4\xi_\da
=-\s0 i4\, \de_{-}(\4\xi^2\4\xi^2{}_\da \4\xi^{\da 1}
-\s0 12\,\4\xi^1\4\xi^2{}_\da \4\xi^{\da 2})
\label{A11b}\eea
one concludes that these monomials are $\de_-$-trivial except for those
which satisfy $s=t=w=0$ and $r<2$. In fact $F$ contains only one
monomial with these properties since \Gl{t13} excludes the values
$r=0$ and $v=0$ in the case $s=t=w=0$ and one concludes
\beq F= \sum_{r\geq 1}(\4\xi^2)^rQ_r(\dh^\a,\4\xi^\da)=\de_-Y
-2k\, \4\xi^2\dh^1.                                  \label{t13j}\eeq
The proof of \Gl{d26} is completed by means of the identity
\bea\4\xi^2\dh^1=\s0 12(\4\xi^2\dh^1-\4\xi^1\dh^2)
+\s0 12(\4\xi^2\dh^1+\4\xi^1\dh^2)
=-\s0 12\, \TH-\s0 i8\, \de_{-}(\4\xi^2{}_\da \4\xi^{\da 1}).
                                                       \label{t13k}\eea

\mysection{Decomposition of $\4\om^{5,\tau}_{nonabel}$}\label{dec}

It is has been mentioned already that \Gl{d8}
does not depend on the gravitino. This remarkable fact holds since the
gravitino depending contributions to $\4\xi^\a$, $\4\xi^\da$ and $\F abi$
cancel in $\4\cF$. Namely evaluating $\4\cF$ explicitly by means of \Gl{t15}
and \Gl{s16} one obtains
\beq \4\cF=\7d\7A+\7A^2+\7L ,\qd \7d=\7C^n\6_n,\qd \7A=\7C^n\A ni T_i,\qd
         \7L=\7C^n(\xi\si_n\La^\dagger-\La\si_n\5\xi)
                                                          \label{d10}\eeq
with $\7C^n$ as in \Gl{t15a}.
Using \Gl{d10} and $\4C=C+\7A$ one easily determines the decomposition
\Gl{d18} of \Gl{d8}:
\beann & &\4\om^{5,\tau}_{nonabel}=\sum_{p=0}^4 \7\om_p,\\
& &\7\om_4 =Tr\{C\7d(\7A\7d\7A+\s0 12\7A^3)
                         +(\7L\7A+\7A\7L)\7d\7A+\s0 32\7L\7A^3
                         -3\, \Xi\, (\xi\La\,\La^\dagger\La^\dagger
                                +\5\xi\La^\dagger\,\La\La)\},\\
& &\7\om_3 =Tr\{-\s0 12\, (C^2\7A+C\7A C+\7A C^2)\7d\7A-\s0 12\, C^2\7A^3
                    +(\7L C+C\7L)\7d\7A\\
        & &\hspace{6ex}
            +\s0 12\, (C\7A^2-\7A C\7A+\7A^2C)\7L+\7A\7L^2\},\\
& &\7\om_2 =\s0 12\, Tr\{-C^3\7d\7A+\7A C\7A C^2
           -(C^2\7A+C\7A C+\7A C^2)\7L+2C\7L^2\},\\
& &\7\om_1 =\s0 12\, Tr(C^4\7A-C^3\7L),\\
& &\7\om_0 =\s0 1{10}\, Tr(C^5).\eeann

\mysection{Useful identities}

\bea & &\Xi=-\s0 1{24}\, \ep_{abcd}\, \4\xi^a\4\xi^b\4\xi^c\4\xi^d
       =-\s0 i{48}\,\4\xi^{\dbe\a}\4\xi_{\a\da}\4\xi^{\da\be}\4\xi_{\be\dbe}
                                                                \label{de1}\\
& &\dh^\a=\4\xi_\da\4\xi^{\da\a},\qd
\5\dh^\da=\4\xi^{\da\a}\4\xi_\a,
\qd \TH=\4\xi_\da\4\xi^{\da\a}\4\xi_\a
                                                           \label{de2a}\\
& &\h^\a=-\s0 i6\, \dh^\be\4\xi_{\be\da}\4\xi^{\da\a},\qd
\5\h^\da=\s0 i{6}\,\4\xi^{\da\a}\4\xi_{\a\dbe}\5\dh^\dbe
                                                              \label{de2b}\\
& &\de_-\Xi=-\s0 13\, \dh^\a\4\xi_{\a\da}\5\dh^{\da},\qd
     \de_-\h^\a=2\, \dh^\a\, \TH=-\dh^\be\dh_\be\4\xi^\a,\qd
     \de_-\5\h^\da=2\, \5\dh^\da\, \TH=\5\dh_\dbe\5\dh^\dbe\4\xi^\da
                                                                \label{de3}\\
& & \de_-(\Xi\, \4\xi^\da)=\s0 12\, \dh^\a\dh_\a\5\dh^\da,\qd
    \de_-(\Xi\, \4\xi^\a)=\s0 12\, \5\dh_\da\5\dh^\da\dh^\a
                                                               \label{de4}\\
& & \4\cd_\da\Xi=-2i\, \5\h_\da=-\s0 i3\, \ep_{abcd}\,
\4\xi^\a\so a\a\da\4\xi^b\4\xi^c\4\xi^d
=-\s0 i3\, \ep_{abcd}\,
\xi^\a\so a\a\da\4\xi^b\4\xi^c\4\xi^d-2i\, \Xi\, \p a\a\so a\a\da
                                                         \label{de5}\\
& &\4\cd_\da\4\cd^\da\Xi=-4i\,\5\dh_\da\5\dh^\da -2M\Xi
                                                               \label{de6}\\
& &\hspace{6em}=8i\4\xi^a\4\xi^b\xi\si_{ab}\xi +16i
\4\xi^a\4\xi^b\4\xi^c\xi\si_{ab}\ps_c
+\Xi\,(16\ps_a\si^{ab}\ps_b-2M)
                                                               \label{de7}\eea

\end{document}